\documentclass[preprint,showpacs,preprintnumbers,amsmath,amssymb]{revtex4}
\usepackage{graphicx}    
\usepackage{dcolumn}     
\usepackage{bm}          
\newcommand{\be}{\begin{equation}}
\newcommand{\ee}{\end{equation}}
\newcommand{\bes}{\begin{equation*}}
\newcommand{\ees}{\end{equation*}}
\newcommand{\bea}{\begin{eqnarray}}
\newcommand{\eea}{\end{eqnarray}}
\newcommand{\beas}{\begin{eqnarray*}}
\newcommand{\eeas}{\end{eqnarray*}}
\newcommand{\ba}{\begin{array}}
\newcommand{\ea}{\end{array}}

\begin{document}
\title{Kondo impurity on the honeycomb lattice at half-filling} 
\author{Luca Dell'Anna}
\affiliation{International School for Advanced Studies, SISSA, I-34014 Trieste, Italy}
\date{\today}
\begin{abstract}
We consider a Kondo-like impurity interacting with fermions on a honeycomb lattice at half-filling, as in the case of graphene.
We derive from the lattice model an effective one-dimensional continuum theory which has, in general, four flavors with angular momentum mixing in the presence of internode scattering processes and six couplings in the spin-isotropic case. Under particular conditions, however, it can be reduced to a single-coupling multichannel pseudogap Kondo model. We finally calculate, in the presence of an energy dependent Fermi velocity, induced by Coulomb interaction, the critical coupling in the large-$N$ expansion, the magnetic susceptibility and the specific heat.
\end{abstract}
\maketitle
\section{Introduction}
\label{intro}
Since the experimental realization of a single monolayer of graphite \cite{novoselov}, named graphene, a two dimensional crystal made of carbon atoms hexagonally packed, a lot of efforts have been made to study many properties of electrons sitting on a honeycomb lattice \cite{castroneto}. Also the problem of magnetic impurities in such a system has become a topic of recent investigations in the last few years \cite{saremi,hentschel,sengpunta,uchoa,cornaglia}, although a detailed derivation of the effective model is still lacking. 

The main motivation of the present work is, therefore, that of deriving, from the lattice Hamiltonian, the corresponding continuum model for the Kondo-like impurity, writing the effective couplings from the lattice parameters. From angular mode expansion we get an effective one-dimensional Kondo model which has, in general, four flavors and is peculiar to graphene-like sublattice systems. Strikingly, we find that there is an angular momentum mixing only in the presence of internode scattering processes, being the valleys and the momenta locked in pairs, in each sublattice sector. The complete model has six couplings in the spin-isotropic case, however, thanks to the lattice symmetry, for some particular positions of the impurity, the number of couplings can be reduced to one, obtaining a multichannel pseudogap Kondo model sharing, now, many similarities with other gapless fermionic systems \cite{withoff,ingersent,ingersent2}, as for example, some semiconductors \cite{withoff}, $d$-wave superconductors \cite{cassanello2} and flux phases \cite{cassanello1}.

A second issue which is worthwhile being addressed is related to interactions. In real systems logarithmic corrections in the density of states may appears, as a result of many-body effects. 
In order to include, at some extent, correlation effects we allow the Fermi velocity to be energy dependent. 
Indeed for a system of electrons in the half-filled honeycomb lattice, like graphene, an effect of Coulomb interaction is that of renormalizing the Fermi velocity \cite{gonzalez94} which grows in the infrared limit. This behavior induces in the density of states subleading logarithmic corrections.
We plan therefore to analyze the effect of these corrections onto the Kondo effect in order to see how finite coupling constant transition, obtained within the large-$N$ expansion technique \cite{read} and renormalization group approach \cite{anderson,hewson}, can be affected by deviations from power law. We find that the critical Kondo coupling becomes non-universal and is enhanced in the ultraviolet by a quantity directly related to the Coulomb screening. 
Moreover, we find that the impurity contribution to the magnetic susceptibility and the specific heat vanish faster by log$^3$ than in the free case, as approaching zero magnetic field or zero temperature. 
\section{The model}
In this section we will derive the continuum one-dimensional effective model from the microscopic lattice Hamiltonian.
\subsection{Lattice Hamiltonian}
\label{sec2}
Let us consider a honeycomb lattice which can be divided into two sublattices, A and B. The tight-binding vectors can be chosen as follows
\bea
&&{\delta}_1=\frac{a}{2}(1,\sqrt{3}),\\
&&{\delta}_2=\frac{a}{2}(1,-\sqrt{3}),\\
&&{\delta}_3=a(-1,0),
\eea
where $a$ is the smallest distance between two sites. These vectors link sites belonging to two different triangular sublattices. Each sublattice is defined by linear combinations of other two vectors, $\frac{a}{2}(3,\sqrt{3})$ and $\frac{a}{2}(3,-\sqrt{3})$. From these values one can derive the reciprocal-lattice vectors in momentum space and draw the Brillouin zone which has an hexagonal shape, i.e. with six corners. We choose two inequivalent corners (the others are obtained by a shift of a reciprocal-lattice vector) at the positions
\bea
\label{K1}
&&{\bf K}=\frac{4\pi}{3\sqrt{3} a}(0,1),\\
\label{K2}
&&{\bf K}^\prime=-{\bf K}.
\eea
These points are actually the Fermi surface reduced to two dots approaching the zero chemical potential, i.e. at half-filling. \\
We will consider the following Hamiltonian defined on this honeycomb lattice
\be
H=H_0+H_{K}.
\ee
The first contribution is given by the tight-binding Hamiltonian
\be
H_0=-t\sum_{{\bf r}\,{\delta}\,\sigma}c_{A\sigma }^\dagger({\bf r}) c_{B\sigma}({\bf r}+{\delta})+h.c.,
\ee
where $t$ is the nearest neighbour hopping parameter, $c_{A\sigma }^\dagger({\bf r})$ ($c_{A\sigma }({\bf r})$) is the creation (annihilation) operator for electrons with spin $\sigma$ localized on the site ${\bf r}$, a vector belonging to the sublattice $A$, while $c_{B\sigma}^\dagger({\bf r}+{\delta})$ ($c_{B\sigma}({\bf r}+{\delta})$) the creation (annihilation) operator for electrons on the site ${\bf r}+\delta$, belonging to the sublattice $B$. 
The second contribution to $H$ is the Kondo-like impurity term
\be
H_{K}=
\sum_{{{\bf v}}}\left\{\lambda_{{{\bf v}} \perp}\left(S_+ c_\uparrow^{\dagger}({\bf v}) c_\downarrow({\bf v})+ S_- c_\downarrow^{\dagger}({\bf v})c_\uparrow({\bf v})\right)
+\lambda_{{{\bf v}} z}S_z \left(c_\uparrow^{\dagger}({\bf v})c_\uparrow({\bf v})-c_\downarrow^{\dagger}({\bf v})c_\downarrow({\bf v})\right)\right\},
\ee
where $\lambda_{{\bf v}\perp}$ and $\lambda_{{\bf v}z}$ are the short-range Kondo couplings, $\vec S$ (with $S_\pm=S_x\pm iS_y$) is the spin of the impurity sitting at the reference position $(0,0)$, $\vec \sigma$ is the spin operator of the electrons located at ${\bf v}$ from the impurity. ${\bf v}$ can belong to $A$ or $B$ and we sum over all these vectors. 
\subsection{Derivation of 1D effective model}
\label{sec3}
We now rewrite the fields $c$ in the following way
\bea
&&c_{A\sigma}({\bf r})\simeq e^{i{\bf K}\cdot {\bf r}}\psi^{{1}}_{A \sigma}{({\bf r})}+e^{-i{\bf K}\cdot {\bf r}}\psi^{{2}}_{A \sigma}{({\bf r})},\\
&&c_{B\sigma}({\bf r}+{\delta})\simeq e^{i{\bf K}\cdot ({\bf r}+{\delta})}\psi^{{1}}_{B\sigma}{({\bf r}+{\delta})}+e^{-i{\bf K}\cdot ({\bf r}+{\delta})}\psi^{{2}}_{B\sigma}{({\bf r}+{\delta})},\\
&&c_{\sigma}({\bf v})\simeq e^{i{\bf K}\cdot {{\bf v}}}\psi^{{1}}_{L\sigma}{({\bf v})}+e^{-i{\bf K}\cdot {{\bf v}}}\psi^{{2}}_{L\sigma}{({\bf v})},
\eea
with $L=A$ if ${\bf v} \in A$, or $L=B$ if ${\bf v} \in B$. The upper indices, $1$ and $2$, label the Fermi points Eqs. (\ref{K1}), (\ref{K2}).  
At these particular points we get the following equalities
\bea
&&\sum_{{\delta}} e^{\pm i{\bf K}\cdot {\delta}}= 0,\\
&&\sum_{{\delta}}{\delta}\, e^{\pm i{\bf K}\cdot {\delta}}= -\frac{3a}{2}(1,\mp i).
\eea
Expanding the slow fields $\psi^i_{L\,\sigma}({\bf r}+{\delta})$ around ${\bf r}$, introducing the multispinor
\be
\label{spinor_spin}
\psi=\left(\psi^{{1}}_{A\,\uparrow},\psi^{{1}}_{B\,\uparrow},\psi^{{2}}_{A\,\uparrow},\psi^{{2}}_{B\,\uparrow},\psi^{{1}}_{A\,\downarrow},\psi^{{1}}_{B\,\downarrow},\psi^{{2}}_{A\,\downarrow},\psi^{{2}}_{B\,\downarrow}\right)^t,
\ee
the identities $\sigma_0$, $\tau_0$, $\gamma_0$ and the Pauli matrices $\sigma_i$, $\tau_{i}$ and $\gamma_{i}$, $i=1,2,3$, acting respectively on the spin space, $\uparrow \downarrow$, valley space, $1, 2$, and sublattice space, $A, B$, we get, in the continuum limit,
\bea
\label{H0}
&&H_0=-i v_F \int\, {d\bf r}\,\psi^\dagger({\bf r})\sigma_0\left(\tau_3 \gamma_1 \partial_y - \tau_0 \gamma_2 \partial_x\right)\psi({\bf r}),\\
&&H_{K}=\psi^\dagger(0)\left(\frac{1}{2}\hat J_{\perp}\left(S_+ \sigma_- + S_- \sigma_+\right)+\hat J_{z}S_z\sigma_z\right)\psi(0),
\eea
where $v_F=\frac{3at}{2}$ is the Fermi velocity and
\be
\label{J_sig}
\hat J_{\sigma}=\frac{1}{2}\left\{\left(J^A_{0\sigma}\tau_0+J^A_{1\sigma}\tau_1+J^A_{2\sigma}\tau_2\right)(\gamma_0+\gamma_3)+\left(J^B_{0\sigma}\tau_0+J^B_{1\sigma}\tau_1+J^B_{2\sigma}\tau_2\right)(\gamma_0-\gamma_3)\right\}
\ee
a Kondo coupling matrix with the following components, containing the lattice details,
\bea
\label{def1}
&&J^L_{0\,\sigma}=
\sum_{{{\bf v}}\in L}\lambda_{{{\bf v}}\,\sigma} \,,\\
\label{def2}
&&J^L_{1\,\sigma}=\sum_{{{\bf v}}\in L}\cos(2{\bf K}\cdot {{\bf v}})\lambda_{{{\bf v}}\,\sigma} \,,\\
\label{def3}
&&J^L_{2\,\sigma}=\sum_{{{\bf v}}\in L}\sin(2{\bf K}\cdot {{\bf v}})\lambda_{{{\bf v}}\,\sigma} \,,
\eea
where $\sigma=\perp, z$ the spin index and $L=A,B$ the sublattice index. Eq.~(\ref{H0}) is a Dirac-Weyl Hamiltonian, constant in spin-space, which, after defining $\bar \psi=\psi^{\dagger}\gamma_3$, can be written as $v_F \int\, {d\bf r}\,\bar \psi ({\bf r})\left(\tau_0 \gamma_1 \partial_x + \tau_3 \gamma_2 \partial_y\right)\psi({\bf r})$ to make Lorentz invariance manifest. 
The spectrum is made of a couple of Dirac cones departing from the two Fermi points, 
and the density of states vanishes linearly approaching the zero energy, $\rho(\epsilon)=\nu |\epsilon|$, where $\nu\propto v_F^{-2}$. This property plays a fundamental role on the scaling behavior of the Kondo impurity, as we are going to see. 
Let us rewrite the full effective Hamiltonian in momentum space, 
\bea
\nonumber H&=&v_F\int\frac{d{\bf p}}{(2\pi)^2}\, \psi^\dagger({\bf p})\,p\,\sigma_0\left(\tau_3 \gamma_1 \sin\theta_p - \tau_0 \gamma_2 \cos\theta_p\right)\psi({\bf p})\\
&&+\int\frac{d{\bf p}}{(2\pi)^2}\int\frac{d{\bf q}}{(2\pi)^2}\, \psi^\dagger({\bf q})\left(\frac{1}{2}\hat J_{\perp}\left(S_+ \sigma_- + S_- \sigma_+\right)+\hat J_{z}S_z\sigma_z\right)\psi({{\bf p}}),
\label{H_moment}
\eea
where we have parametrized the momenta as follows
\bea
p_x=p\,\cos\theta_p ,\\
p_y=p\,\sin\theta_p .
\eea
For the benefits of forthcoming discussions we first notice that 
the orbital angular momentum operator 
\be
{\cal L}=-i(x\partial_y-y\partial_x)
\ee 
does not commute with the Hamiltonian $H_0$ in Eq.~(\ref{H0}). 
On the other hand, in order to define proper total angular momenta we introduce the operator 
\be
{\cal J}={\cal L}+\frac{1}{2}\tau_3\gamma_3,
\ee
which does commute with $H_0$,
\be
\left[{\cal J},H_0\right]=0,
\ee
and also with the $\tau_0$-components of $H_K$. In particular, given some amplitudes $\psi^{{i}}_{L\sigma m}(p)$, an eigenstate of ${\cal J}$ with eigenvalue $j=\left(m+\frac{1}{2}\right)$ can be written as
\be
\label{eigenJ}
\left(\ba{r}
e^{im\theta_p} \psi^{{1}}_{A\,\sigma\, m}(p)\\
e^{i(m+1)\theta_p} \psi^{{1}}_{B\,\sigma\, m}(p)\\
e^{i(m+1)\theta_p} \psi^{{2}}_{A\,\sigma\, m}(p)\\
e^{im\theta_p}\psi^{{2}}_{B\,\sigma\, m}(p)
\ea\right).
\ee
Performing the following unitary transformation 
\be
\label{U}
U_p= \frac{e^{\frac{i}{2}\theta_p\tau_3}}{2\sqrt{2}}\tau_0\left[(1+ie^{-i\theta_p\tau_3})(\gamma_0-i\gamma_2)+(1-ie^{-i\theta_p\tau_3})(\gamma_1-\gamma_3)\right]
\ee
to the fields
\be
\label{Uphi}
 \psi({\bf p})=U_p \,\phi({\bf p}),
\ee
the Hamiltonian Eq.~(\ref{H_moment}) becomes
\bea
\label{H_rot}
 H&=&v_F\int\frac{d{\bf p}}{(2\pi)^2}\, \phi^\dagger({\bf p})\,p\,\sigma_0\tau_0\gamma_3\,\phi({\bf p})\\
\nonumber &&+\int\frac{d{\bf p}}{(2\pi)^2}\int\frac{d{\bf q}}{(2\pi)^2}\, \phi^\dagger({\bf q})\left(\frac{1}{2}\hat K_{\perp}(\theta_q,\theta_p)\left(S_+ \sigma_- + S_- \sigma_+\right)+\hat K_{z}(\theta_q,\theta_p)S_z\sigma_z\right)\phi({{\bf p}}),
\eea
namely, $H_0$ becomes diagonal, the cost to pay is that the Kondo couplings depends on the angular part of the momenta,
\bea
\label{K_rot}
\nonumber \hat K_{\sigma}(\theta_q,\theta_p)&\equiv& \frac{1}{2}\Big\{
\left(J_{0\sigma}^{A} e^{\frac{i}{2}(\theta_q-\theta_p)\tau_3}\tau_0 +J_{1\sigma}^{A}e^{\frac{i}{2}(\theta_q+\theta_p)\tau_3}\tau_1+J_{2\sigma}^{A}e^{\frac{i}{2}(\theta_q+\theta_p)\tau_3}\tau_2\right)(\gamma_0-\gamma_1)\\
&&+
\left(J_{0\sigma}^{B} e^{\frac{i}{2}(\theta_p-\theta_q)\tau_3}\tau_0+J_{1\sigma}^{B} e^{-\frac{i}{2}(\theta_q+\theta_p)\tau_3}\tau_1+J_{2\sigma}^{B} e^{-\frac{i}{2}(\theta_q+\theta_p)\tau_3}\tau_2\right)(\gamma_0+\gamma_1)
\Big\}.
\eea 
Notice that the angular dependence of $\hat K$ does not prevent the model to be renormalizable. As one can see by poor man's scaling procedure \cite{anderson,hewson}, being ${\bf q}$ the intermediate momentum in the edge bands, dropping for the moment the spin indices, the contributions which renormalize, for instance, $J_{0}^{A}$ in the particle channel are
\bea
&& J_{0}^{A}J_{0}^{A} e^{\frac{i}{2}(\theta_{p_1}-\theta_{q})\tau_3}e^{\frac{i}{2}(\theta_{q}-\theta_{p_2})\tau_3}[...]=J_{0}^{A}J_{0}^{A} e^{\frac{i}{2}(\theta_{p_1}-\theta_{p_2})\tau_3}[...],\\
&&J_{i}^{A}J_{i}^{A} e^{\frac{i}{2}(\theta_{p_1}+\theta_{q})\tau_3}\tau_i e^{\frac{i}{2}(\theta_{p_2}+\theta_{q})\tau_3}\tau_i[...]=J_{i}^{A}J_{i}^{A}  e^{\frac{i}{2}(\theta_{p_1}-\theta_{p_2})\tau_3}[...]
\eea 
with $i=1,2$. 
In the same way we can check that the corrections to $J_{i}^{A}$, with $i=1,2$, are
\bea
J_{0}^{A}J_{i}^{A}  e^{\frac{i}{2}(\theta_{p_1}-\theta_{q})\tau_3}e^{\frac{i}{2}(\theta_{q}+\theta_{p_2})\tau_3}\tau_i[...]= J_{0}^{A}J_{i}^{A} e^{\frac{i}{2}(\theta_{p_1}+\theta_{p_2})\tau_3}\tau_i[...],\\
J_{i}^{A}J_{0}^{A}   e^{\frac{i}{2}(\theta_{p_1}+\theta_{q})\tau_3}\tau_i e^{\frac{i}{2}(\theta_{q}-\theta_{p_2})\tau_3}[...]=J_{i}^{A}J_{0}^{A}  e^{\frac{i}{2}(\theta_{p_1}+\theta_{p_2})\tau_3}\tau_i[...].
\eea
Analogous corrections can be verified in the hole channel. In all these corrections  $\theta_q$ always cancels out, recovering the right momentum dependence for the slow modes.\\ 
From Eqs.~(\ref{def1}-\ref{def3}) we actually get access to the renormalization of linear combinations of the original lattice parameters $\lambda_{{\bf v}}$. 

In order to reduce the problem to one dimension we proceed expanding the fields $\phi({\bf p})$ in angular momentum eigenmodes as follows
\be
\phi({\bf p})=\sum_{m=\infty}^\infty e^{i (m+\frac{1}{2})\theta_p}\phi_m(p),
\ee
with $m\in \mathbb{Z}$. Indeed, due to the gauge in Eq.~(\ref{U}), all the spinor components have the same angular phase. Actually from Eq.~(\ref{Uphi}), one verify that $e^{i (m+\frac{1}{2})\theta_p}\phi_m(p)$ is the eigenvector of ${\cal J}$, Eq.~(\ref{eigenJ}), with eigenvalue $j=m+\frac{1}{2}$, transformed by $U_p^{-1}$, and with amplitudes 
\be
\phi^{{i}}_{\pm\,\sigma \,m}(p) =\frac{1}{\sqrt{2}}\left(\psi^{{i}}_{B\,\sigma\, m}(p)\mp i\psi^{{i}}_{A\,\sigma\, m}(p)\right),
\ee
where the subscript $\pm$ replaces the sublattice index and refers to the sign of the energy, $v_Fp\gamma_3$, appearing in Eq.~(\ref{H_rot}). 
The original field at position ${\bf r}=(r,\varphi)$, can be written as 
\be
\psi(r,\varphi)=\int \frac{dp\,p}{4\sqrt{2}\pi}\sum_{m=-\infty}^{\infty}i^me^{im\varphi}
\left(\ba{c}
i\left[{\sf J}_{m+1}(pr)e^{i\varphi}\gamma_d+
{\sf J}_m(pr)\gamma_u\right]\phi^{{1}}_{m}(p)\\
\left[{\sf J}_{m}(pr)\gamma_d-
{\sf J}_{m+1}(pr)e^{i\varphi}\gamma_u\right]\phi^{{2}}_{m}(p)
\ea
\right),
\ee
where ${\sf J}_m(z)$ are the Bessel functions of the first kind, $\gamma_d\equiv\gamma_0+\gamma_1-i\gamma_2-\gamma_3$ and $\gamma_u\equiv\gamma_0-\gamma_1-i\gamma_2+\gamma_3$. At $r=0$ the only terms which survive are those with $m=0,-1$, corresponding to $j=\pm \frac{1}{2}$, in terms of eigenvalues of ${\cal J}$. 
After integrating Eq.~(\ref{H_rot}) over the angles, indeed, we get in the $H_K$ only contributions with $m=0,-1$, in the following combinations 
\bea
\nonumber H &=&\int_0^{\infty} dp\, \frac{v_Fp^2}{2\pi} \sum_{i=1,2} \sum_{m=-\infty}^{\infty} \phi^{{i}\dagger}_{m}(p)\,\sigma_0\gamma_3\,\phi^{{i}}_{m}(p)  
+
\frac{1}{2}
\int_0^{\infty}\frac{dp}{2\pi}\,p\int_0^{\infty}\frac{dq}{2\pi}\, q\\
\nonumber&& \Big\{
J^{A}_{0}\left[\phi^{{1}\dagger}_{0}(q) 
({\vec S}\cdot{\vec \sigma})
(\gamma_0-\gamma_1)\phi^{{1}}_{0}(p)
+\phi^{{2}\dagger}_{-1}(q) 
({\vec S}\cdot{\vec \sigma})
(\gamma_0-\gamma_1)\phi^{{2}}_{-1}(p)\right]\\
\nonumber &&+J^{A}_{1}\left[\phi^{{1}\dagger}_{0}(q) 
({\vec S}\cdot{\vec \sigma})
(\gamma_0-\gamma_1)\phi^{{2}}_{-1}(p)
+\phi^{{2}\dagger}_{-1}(q) 
({\vec S}\cdot{\vec \sigma})
(\gamma_0-\gamma_1)\phi^{{1}}_{0}(p)\right]\\
\nonumber &&-i J^{A}_{2}\left[\phi^{{1}\dagger}_{0}(q) 
({\vec S}\cdot{\vec \sigma})
(\gamma_0-\gamma_1)\phi^{{2}}_{-1}(p)
-\phi^{{2}\dagger}_{-1}(q) 
({\vec S}\cdot{\vec \sigma})
(\gamma_0-\gamma_1)\phi^{{1}}_{0}(p)\right]
%
\\
\nonumber&&+
J^{B}_{0}\left[\phi^{{1}\dagger}_{-1}(q) ({\vec S}\cdot{\vec \sigma})(\gamma_0+\gamma_1)\phi^{{1}}_{-1}(p)+\phi^{{2}\dagger}_{0}(q) 
({\vec S}\cdot{\vec \sigma})(\gamma_0+\gamma_1)\phi^{{2}}_{0}(p) \right]\\
\nonumber &&+J^{B}_{1}\left[\phi^{{1}\dagger}_{-1}(q) 
({\vec S}\cdot{\vec \sigma})
(\gamma_0+\gamma_1)\phi^{{2}}_{0}(p) 
+\phi^{{2}\dagger}_{0}(q) ({\vec S}\cdot{\vec \sigma})
(\gamma_0+\gamma_1)\phi^{{1}}_{-1}(p) \right]\\
 &&-i J^{B}_{2}\left[\phi^{{1}\dagger}_{-1}(q) 
({\vec S}\cdot{\vec \sigma})
(\gamma_0+\gamma_1)\phi^{{2}}_{0}(p)
-\phi^{{2}\dagger}_{0}(q) 
({\vec S}\cdot{\vec \sigma})
(\gamma_0+\gamma_1)\phi^{{1}}_{-1}(p)\right]
\Big\},
\eea
where now $\phi^{{i}}_{m}(p)$ are spinors only in spin and energy spaces. Here we are considering the spin-isotropic case, with $J^L_{i}\equiv J^L_{i\perp}=J^L_{i z}$, to simplify the notation. In the spin-anisotropic case one simply has to replace $J^L_{i}({\vec S}\cdot{\vec \sigma})$ with $J^{L}_{i\perp}({S}_x {\sigma}_x+{S}_y {\sigma}_y)+J^{L}_{i z}({S}_z{\sigma}_z)$. 
In the free part of the effective model, $H_0$, we keep only the contributions from the particles with $m=0,-1$, the only ones which can scatter with the impurity. We now unfold the momenta from $[0,+\infty)$ to $(-\infty,+\infty)$ by redefining the fields in the following way 
\be
\xi^{{i}}_{s\,\sigma}(p)\equiv \left[{\textrm {sign}}(p)\right]^{m+i}\sqrt{|p|}\phi^{{i}}_{{\textrm {sign}}(p)\,\sigma\, m}(|p|),
\ee
where, in order to label the fermions, we choose the index $i$ in valley space and the index $s={\textrm {sign}}\left(m+\frac{1}{2}\right)$, the sign of the total angular momenta, eigenvalues of ${\cal J}$, which are good quantum numbers as soon as there is not internode scattering, i.e. $J^{A}_1=J^{B}_1=J^{A}_2=J^{B}_2=0$. Introducing for simplicity  
\be
J^{L}_{\pm}\equiv J^{L}_{1}\pm i J^{L}_{2} =\sum_{{{\bf v}}\in {L}}e^{\pm i 2 {\bf K}\cdot {{\bf v}}}\lambda_{{{\bf v}}} \,,
\ee
we finally end up with the following one-dimensional effective Hamiltonian
\bea
\label{H_full}
H&=&\int_{-\infty}^{\infty} \frac{dp}{2\pi}\,E(p)\sum_{s,\sigma,i}\xi^{{i}\dagger}_{s\sigma}(p)\,\xi^{{i}}_{s\sigma}(p)
+\frac{1}{2}\int_{-\infty}^{\infty} \frac{dq}{2\pi}\,\int_{-\infty}^{\infty} \frac{dp}{2\pi}\sqrt{|q|}\sqrt{|p|}\\
\nonumber&&{\vec S}\cdot\Big\{
J^{A}_{0}\left(\xi^{{1}\dagger}_{{+}}(q) 
{\vec \sigma}
\xi^{{1}}_{{+}}(p)
+\xi^{{2}\dagger}_{{-}}(q) {\vec \sigma}
\xi^{{2}}_{{-}}(p)\right)
%
+J^{A}_{-}\,\xi^{{1}\dagger}_{{+}}(q) 
{\vec \sigma}
\xi^{{2}}_{{-}}(p)
+J^{A}_{+}\,\xi^{{2}\dagger}_{{-}}(q) 
{\vec \sigma}
\xi^{{1}}_{{+}}(p)\\
\nonumber &&\phantom {S}+J^{B}_{0}\left(\xi^{{1}\dagger}_{{-}}(q) 
{\vec \sigma}\xi^{{1}}_{{-}}(p)
+\xi^{{2}\dagger}_{{+}}(q) {\vec \sigma}\xi^{{2}}_{{+}}(p)\right)
+J^{B}_{-} \, \xi^{{1}\dagger}_{{-}}(q) 
{\vec \sigma}
\xi^{{2}}_{{+}}(p)
+J^{B}_{+} \, \xi^{{2}\dagger}_{{+}}(q) 
{\vec \sigma}
\xi^{{1}}_{{-}}(p)
\Big\},
\eea
where, in the first term, the indices $s=\pm$, $\,i=1,2$ and $\sigma=\uparrow,\downarrow$, are summed, and the dispersion relation is $E(p)=v_F p$.
The full model, Eq.~(\ref{H_full}), has six Kondo couplings, in the spin-isotropic case, which are independent for a generic position of the magnetic impurity on the lattice. Moreover Eq.~(\ref{H_full}) exhibits an angular momentum mixing in the presence of internode scattering amplitudes $J^A_\pm$ and $J^B_\pm$, namely, when also the nodes are mixed. We are not going to analyze the complete model in full generality but we shall consider only particular cases physically relevant. 

\subsection{Some particular examples}

\paragraph{Impurity on a site.} 
If we consider an impurity on top of a site of the honeycomb lattice, belonging to the sublattice $A$, for instance, and consider only the 
nearest neighbour coupling between the impurity and the electrons located on this site, we have $\lambda_{{\bf v}}\neq 0$ if ${\bf v}=(0,0)$ and assume $\lambda_{{\bf v}}=0$ for ${\bf v}\neq (0,0)$. In this case 
we get 
\bea
&&J^A_{0}=J^A_{1}=\lambda_{(0,0)}, \\
&&J^A_{2}=J^B_{0}=J^B_{1}=J^B_{2}=0.
\eea
Introducing the symmetric combination for the fields 
\be
\zeta=\xi^{{1}}_{+}+\xi^{{2}}_{-} \,,
\ee
the effective Hamiltonian Eq.~(\ref{H_full}) becomes simply
\be
H=\int_{-\infty}^{\infty} \frac{dp}{2\pi}\,E(p)\sum_{\sigma}\zeta^{\dagger}_{\sigma}(p)\zeta_{\sigma}(p)
+\frac{J^{A}_{0}}{2}\iint_{-\infty}^{\infty} \frac{dq}{2\pi} \frac{dp}{2\pi}{\sqrt{|qp|}}\,{\vec S}\cdot \zeta^{\dagger}(q) {\vec \sigma}\zeta(p) ,
\label{H_os}
\ee
which is a single channel Kondo model.

\paragraph{Impurity by substitution.} 
If we now consider an impurity sitting on a site of the honeycomb lattice, let us say, belonging to the sublattice $A$, and consider only 
nearest neighbour couplings between the impurity and the electrons, we have $\lambda_{{\bf v}}=0$ if ${\bf v} \in A$ while $\lambda_{{\bf v}}=\lambda_{{\delta}_1}=\lambda_{{\delta}_2}=\lambda_{{\delta}_3}$, if ${\bf v} = {\delta}_i$, $i=1,2,3$, and $\lambda_{{\bf v}}=0$ for $v>a$. Noticing that
\be
\label{key}
\sum_{i=1}^3\cos(2{\bf K}\cdot {\delta}_i)=\sum_{i=1}^3 \sin(2{\bf K}\cdot {\delta}_i)=0,
\ee 
we get a remarkable reduction of the number of couplings given by Eqs.~(\ref{def1}-\ref{def3})
\be
J^A_{0}=J^A_{1}=J^A_{2}=J^B_{1}=J^B_{2}=0.
\ee
Recalling the fields as follows
\be
\zeta_{1}=\xi^{{1}}_{-}\,,\;\;\;
\zeta_{2}=\xi^{{2}}_{+} \,,
\ee
the effective Hamiltonian Eq.~(\ref{H_full}) reduces to 
\be
H=\int_{-\infty}^{\infty} \frac{dp}{2\pi}\,E(p)\sum_{\sigma,\,i}\zeta^{\dagger}_{i\sigma}(p)\zeta_{i\sigma}(p)+\frac{J^{B}_{0}}{2}\iint_{-\infty}^{\infty} \frac{dq}{2\pi} \frac{dp}{2\pi}{\sqrt{|qp|}}\sum_{i=1}^{N_f}{\vec S}\cdot\zeta^{\dagger}_{i}(q) {\vec \sigma}\zeta_{i}(p) ,
\label{H_sub}
\ee
where the $N_f=2$ flavors (the valleys and the momenta are locked in pairs) are decoupled and we realize a two-channel Kondo model. The reduced model Eq.~(\ref{H_sub}) is the same as that found for flux phases \cite{cassanello1}.

\paragraph{Impurity at the center of the cell.} 
Finally, let us consider an impurity at the center of the honeycomb cell. In this case, using Eqs.~(\ref{def1}-\ref{def3}) and Eq.~(\ref{key}), we have
\bea
&&J^A_{0}=J^B_{0},\\
&&J^A_{1}=J^A_{2}=J^B_{1}=J^B_{2}=0.
\eea
Enumerating the fields as follows, for instance,
\be
\zeta_{1}=\xi^{{1}}_{+}\,,\;\;\; 
\zeta_{2}=\xi^{{2}}_{-}\,,\;\;\;
\zeta_{3}=\xi^{{1}}_{-}\,,\;\;\;
\zeta_{4}=\xi^{{2}}_{+}\,,
\ee
we get the same Hamiltonian as in Eq.~(\ref{H_sub}) with, now, $N_f=4$ flavors, realizing, therefore, a four-channel Kondo model, as in the case of $d$-wave superconductors \cite{cassanello2}.
\section{Large-$N$ expansion and the role of Coulomb interaction}
In this section we solve the model Eq.~(\ref{H_sub}) in the large-$N$ approximation, where $N$ is the rank of the symmetry group of the impurity, which actually is equal to $2$ for spin one-half. Following the standard procedure \cite{read, cassanello2}, within a path integral formalism, we write $\vec S=f_\alpha^{\dagger}\vec \sigma_{\alpha\beta} f_\beta$, introducing additional fermionic fields $f$, with the constraint $Q=f_\alpha^{\dagger} f_\alpha$, the charge occupancy at the impurity site. In the Lagrangian, therefore, a Lagrange multiplier $\epsilon_0$ is included to enforce such constraint, which is actually the impurity Fermi level. To decouple the quartic fermionic term one introduces the Hubbard-Stratonovich fields $\Phi_i$, where $i=1,... N_f$, being $N_f$ the number of flavors. For impurity by substitution $N_f=2$, as seen before. After integrating over the fermionic fields, $\zeta$ and $f$, we end up with the following effective free energy,
\be
F=\frac{N}{\pi} \int d\epsilon\, f(\epsilon)\,\delta(\epsilon)+\int d\tau \left(\frac{N}{J^B_0}\sum_i^{N_f} |\Phi_i(\tau)|^2-Q\,\epsilon_0\right),
\label{F}
\ee
where $f(\epsilon)$ is the Fermi function and 
\be
\delta(\epsilon)=\arctan\left(\frac{\pi|\epsilon|\Delta/2}{(\epsilon-\epsilon_0)v_F^2+\epsilon\Delta \ln(\Lambda/|\epsilon|)}\right) 
\ee
the phase shift, with $\Delta=\sum_i |\Phi_i(\epsilon)|^2/\pi$, and $\Lambda$ a positive ultraviolet cut-off which dictates the limit of validity of the continuum Dirac-like model for the free Hamiltonian. For graphene the typical value is $\Lambda\sim 2$eV.

So far we have considered a model of free fermions hopping on a lattice and scattering eventually with a magnetic impurity, but in order to get more realistic predictions we should consider, at some extent, interaction effects. 
In order to do that, we let the Fermi velocity be energy dependent, i.e. $v_F\equiv v_F(\epsilon)$. 

This is not unrealistic since it has been shown \cite{gonzalez94} that, due to Coulomb screening in an electronic system defined on the half-filled honeycomb lattice, as in the case of a monolayer of graphene \cite{gonzalez99,polini}, the effective Fermi velocity is renormalized in such a way that $v_F$ flows to higher values in the infrared, and consequently the density of states around the Fermi energy decreases. The low energy behavior for the renormalized velocity is $v_F\sim \ln(\epsilon^{-1})$ and so the density of states should behave naively as $\rho\sim \epsilon\, v_F^{-2}\sim \epsilon/\ln(\epsilon^{-1})^2$. 
The aim of the following section is then to study the role of such corrections onto the Kondo effect, neglecting, however, possible renormalization of the Kondo coupling due to Coulomb interaction. The idea is to consider an uncharged magnetic impurity embedded in a cloud of charges dressed by Coulomb interaction. The realistic expression for the Fermi velocity is the following \cite{polini}
\be
\label{vF}
v_F(\epsilon)= v\left(1+\eta \ln(\Lambda/|\epsilon|)\right),
\ee
where $\eta$ is related to the fine structure constant, for Thomas-Fermi screening it is $\eta=e^2/4\varepsilon \hbar v$, being $\varepsilon$ the dielectric constant, and $v$ is an energy independent velocity.

\subsection{Saddle point equations}

From Eq.~(\ref{F}), the extremal values of $\epsilon_0$ and $\Delta$, evaluated at zero energy in the static approximation, satisfy the saddle point equations $\frac{\partial F}{\partial \epsilon_0}=0$ and $\frac{\partial F}{\partial \Delta}=0$, which can be written as follows \cite{cassanello1,cassanello2}
\bea
\label{Qsp}
Q&=&\frac{1}{\pi}\int^{D}_{-D}d\epsilon \, f(\epsilon) \frac{\partial\delta}{\partial \epsilon_0}(\epsilon),\\
-\frac{1}{J^B_0}&=&\frac{1}{\pi^2}\int^{D}_{-D} d\epsilon \, f(\epsilon) \frac{\partial\delta}{\partial \Delta}(\epsilon),
\label{Jsp}
\eea
where $D\le \Lambda$ is the bandwidth. The Eq.~(\ref{Qsp}) dictates the relation between the singlet amplitude $\Delta\sim\sum_i\langle|\sum_{\sigma}\zeta_{i\sigma}^\dagger f_{\sigma}|^2\rangle$ and the impurity level $\epsilon_0$, at fixed occupation charge $Q$, and reads
\be
\label{Q}
Q=\int_{-D}^D \frac{d\epsilon}{\pi}\, f(\epsilon) \frac{2\pi v_F(\epsilon)^2|\epsilon|\Delta}{(\pi|\epsilon|\Delta)^2+4(v_F(\epsilon)^2(\epsilon-\epsilon_0)+\epsilon\Delta\ln(\Lambda/|\epsilon|))^2}.
\ee
For $T=0$ and for a generic value of $Q$, we get the following behavior for the impurity level, 
$\epsilon_0\sim \Lambda\, e^{\frac{1}{\eta}\left(1+\frac{\Delta (1+\eta\ln(\Lambda/D))}{2\eta Qv^2(1+\eta\ln(\Lambda/D))-\Delta}\right)}$. In the non-interacting limit, formally, when $\eta\rightarrow 0$, it reduces to $\epsilon_0\sim D\, e^{-{2v^2 Q}/{\Delta}} $, in agreement with Ref.~\cite{cassanello2}. Strikingly, the limit of $\Delta\rightarrow 0$ is finite and equal to $\Lambda\,e^{1/\eta}$, i.e. the two limits do not commute. This means that, in the presence of Coulomb interaction, the occupation charge for an impurity level within the bandwidth is finite only if the singlet is formed and $Q\sim \Delta/2\eta v v_F(D)$. The energy scale $\epsilon_0$ in Eq.~(\ref{Q}) does not play the role of an infrared cut-off for $\Delta\rightarrow 0$, and as a result, in that limit, $Q$ goes to zero for any value of $\epsilon_0$. This result is different from that found in the free case \cite{cassanello2} where the Fermi velocity is constant, $v_F(\epsilon)=v$. In the latter case $\epsilon_0$ vanishes, approaching zero singlet amplitude, for any value of the occupation charge.

The second equation, Eq.~(\ref{Jsp}), dropping the indices for simplicity, reads
\be
\frac{1}{J}=\int_{-D}^D \frac{d\epsilon}{\pi}\, f(\epsilon) \frac{2v_F(\epsilon)^2|\epsilon|(\epsilon_0-\epsilon)}{(\pi|\epsilon|\Delta)^2+4(v_F(\epsilon)^2(\epsilon-\epsilon_0)+\epsilon\Delta\ln(\Lambda/|\epsilon|))^2}.
\ee
Setting $\epsilon_0=0$ and $\Delta=0$ at $T=0$, we get the following critical coupling
\be
\label{Jc}
\frac{1}{J_c}=\int_{0}^D\frac{d\epsilon}{2\pi} \frac{1}{v_F(\epsilon)^2}=\frac{D}{2\pi\eta^2v^2}\left\{\frac{\eta }{1+\eta \ln(\Lambda/D)}-\frac{\Lambda}{D} e^{1/\eta}\Gamma[0,1/\eta+\ln(\Lambda/D)]\right\},
\ee 
where $\Gamma[a,x]\equiv \int^\infty_x t^{a-1}e^{-t}dt$  
is the Incomplete Gamma function. Sending $\eta\rightarrow 0$ we recover the standard result $\frac{1}{J_c}=\frac{D}{2\pi v^2}$ \cite{withoff}.\\ 
At this point it is worthwhile making a digression. Contrary to the free case, where the limit $\lim_{D\rightarrow 0}\frac{v^2}{DJ_c}$ is trivially finite and equal to $\frac{1}{2\pi}$, in the interacting case, using Eq.~(\ref{Jc}), this limit is zero. On the other hand, if we replace $v$ with renormalized velocity $v_F(D)$, the limit
\be
\lim_{D\rightarrow 0}\frac{v_F(D)^2}{DJ_c}=\frac{1}{2\pi}
\ee
is finite and equal to the standard case. This is consistent with the fact that  the dimensionless parameter relevant in the Kondo effect is not the bare coupling $J$ but the product $\rho J$ and that, in the presence of a renormalized Fermi velocity, Eq.~(\ref{vF}), the density of states is modified as $\rho(\epsilon)\sim\epsilon/v_F(\epsilon)^2$. In order to validate this result and to get more insights one can address the problem from a renormalization group prospective, as we did in Appendix.\\
From Eq.~(\ref{Jc}) we find that the critical coupling $J_c\rho(D)$ is not universal, being an increasing function of the ratio $D/\Lambda$, and is larger than the corresponding mean field result in the non-interacting case for any positive $D\le \Lambda$.\\To go beyond the tree level, one should consider quantum fluctuations, i.e. higher orders in large-$N$ expansion, which might spoil the critical point obtained in the mean field level, as in the case of strictly power-law pseudogap Kondo systems \cite{ingersent2}, if the particle-hole symmetry is preserved. In order to break particle-hole symmetry, however, one can include straightforwardly a gate voltage in the model \cite{sengpunta}. In any case the role of fluctuations, in the presence of a logarithmic deviation from power-law in the density of states is still an open issue which we are not going to address here. 

\subsection{Magnetic susceptibility and specific heat}
\paragraph{Magnetic susceptibility.} 
The magnetic field can be easily included in our final model introducing a Zeeman term $H\sigma_3$. This term modifies the phase shift in the free energy, Eq.~(\ref{F}), as $\delta(\epsilon)\rightarrow \frac{1}{2}(\delta(\epsilon+H)+\delta(\epsilon-H))$. We can, therefore, calculate the magnetization 
\be
M(T,H)=-\frac{\partial F}{\partial H}, 
\ee
and the magnetic susceptibility 
\be
\chi(T,H)=-\frac{\partial^2 F}{\partial H^2}.
\ee
For $T\rightarrow 0$ and $H\ll \epsilon_0$, we have the following magnetization
\be
M(0,H)=\frac{N}{2\pi}\left[\delta(-H)-\delta(H)\right]\simeq \frac{N\Delta H^2}{2\epsilon_0^2 v_F(H)^4}\left[v_F(H)^2+\Delta\ln(\Lambda/H)\right]\simeq \frac{N\Delta}{2\epsilon_0^2 v^2\eta^2}\frac{H^2}{\ln(\Lambda/H)^2}.
\ee
The final result for the magnetization is valid only if 
$H\ll \Lambda e^{-\Delta/(v\eta)^2}$. 
In the same limit the asymptotic behavior of the magnetic susceptibility is, then, given by
\be
\label{chi(0,H)}
\chi(0,H)\simeq \frac{N\Delta}{\epsilon_0^2 v^2\eta^2}
\frac{H}{\ln(\Lambda/H)^2}.
\ee
For $\Lambda e^{-\Delta/(\eta v)^2}\ll H\ll\epsilon_0$, instead, one gets $\chi(0,H)\simeq \frac{N\Delta^2 H}{\epsilon_0^2 v_F(H)^4}\ln({\Lambda}/{H})$, and for $v_F(H)\rightarrow v$, one recover the result for the non-interacting case \cite{cassanello2}.\\
For $H\rightarrow 0$ and $T\ll \epsilon_0,\,\Lambda e^{-\Delta/(\eta v)^2}$, we have, instead, the following magnetic susceptibility
\be
\chi(T,0)=\frac{N}{\pi}\int_{-\infty}^{\infty} d\epsilon \frac{\partial f}{\partial\epsilon}\frac{\partial\delta}{\partial\epsilon}\simeq \frac{N\Delta T}{2\epsilon_0^2}\int_{-\infty}^{\infty} dx \frac{e^{x}(e^x-1)}{(1+e^x)^3}\frac{|x|x}{v_F(T|x|)^2}
\simeq \frac{2\ln(2)N\Delta}{\epsilon_0^2 v^2\eta^2}\frac{T}{\ln(\Lambda/T)^2},
\ee
which crosses over to $\chi(T,0)=\frac{2\ln(2)N\Delta^2}{\epsilon_0^2 v_F(T)^4} T\,{\ln(\Lambda/T)}$, for $\Lambda e^{-\Delta/(\eta v)^2}\ll T\ll \epsilon_0$.
In the presence of Coulomb interaction, therefore, the magnetization and the susceptibility vanish logarithmically faster, as approaching zero magnetic field or zero temperature, than in the free pseudogap case. The overscreening effects pointed out in Ref.~\cite{cassanello2} is, then, enhanced by a factor on the order $\frac{\Delta\eta^2}{v^2}[\ln(\frac{\Lambda}{{\textrm{min}}(T,H)})]^3$, in the presence of a renormalized Fermi velocity induced by the Coulomb screening. 
\paragraph{Specific heat.} Let us calculate now the impurity contribution to the specific heat, defined as follows 
\be
C(T,H)=-T\frac{\partial F}{\partial T^2}.
\ee
For $H\rightarrow 0$ and $T\ll \epsilon_0$, we have
\bea
\nonumber C(T,0)=\frac{N}{T\pi}\int_{-\infty}^{\infty} d\epsilon \,\epsilon^2\frac{\partial f}{\partial\epsilon}\frac{\partial\delta}{\partial\epsilon}
\simeq \frac{N\Delta T^2}{2\epsilon_0^2}\int_{-\infty}^{\infty} dx \left(\frac{x^2e^{x}(e^x-1)}{(1+e^x)^3}-\frac{2xe^x}{(1+e^x)^2}\right)\frac{|x|x}{v_F(T|x|)^2}\\
\simeq 
\frac{9\zeta(3)N\Delta}{\epsilon_0^2 v^2\eta^2}\frac{T^2}{\ln(\Lambda/T)^2},\;
\label{C(T,0)}
\eea
where $\zeta(3)\approx 1.2$ is the Riemann zeta function at $3$. Also in this case we assume $T\ll \Lambda e^{-\Delta/(\eta v)^2}$. For $T$ larger than that energy scale, instead, $C(T,0)\simeq \frac{9\zeta(3)N\Delta^2}{\epsilon_0^2 v_F(T)^4}{T^2}{\ln(\Lambda/T)}$, and for $\eta\rightarrow 0$ one recover the non-interacting result \cite{cassanello2}.\\
For $T\rightarrow 0$ and $H\ll \epsilon_0$ we have, instead, the following behavior 
\bea
\nonumber C(T,H)=\frac{N}{2\pi T}\int_{-\infty}^{\infty} d\epsilon \,\epsilon^2 \frac{\partial f}{\partial\epsilon}\left(\frac{\partial \delta}{\partial\epsilon}(\epsilon+H)+\frac{\partial \delta}{\partial\epsilon}(\epsilon-H)\right)\simeq -T\chi(0,H)\int_{-\infty}^{\infty} dx x^2\frac{\partial f}{\partial x}\\
=\frac{\pi^2}{3}T \chi(0,H),
\label{C(0,H)}
\eea
with $\chi(0,H)$ calculated before. Eq.~(\ref{C(0,H)}) corresponds also to the asyptotic behavior of the impurity entropy, $-\partial F/\partial T$, 
for $T\ll H$.
The specific heat, like the magnetic susceptibility, vanishes logarithmically faster than in the case with constant Fermi velocity, approaching zero magnetic field and zero temperature. 
However, the Wilson ratios $C/(T \chi)$, both for $T\ll H$ and for $H\ll T$, are exactly the same as those found in Ref.~\cite{cassanello2}. 

\section{Summary and conclusions}
We have derived the low-energy continuum limit of a Kondo-like impurity model defined on a honeycomb lattice at half-filling. 

By angular momentum eigenmode expansion we have obtained an effective one-dimensional model with two colors and four flavors, two for each sublattice sector, Eq.~(\ref{H_full}). The impurity effective Hamiltonian involves two angular momenta which are linked with the two nodes. We have found, therefore, that the internode scattering contributions correspond also to the angular momentum mixing terms. Quite in general, we have to deal with six couplings in the spin-isotropic case, which are linear combinations of the original lattice parameters, Eqs.~(\ref{def1}-\ref{def3}). However, due to the underlying lattice symmetry, in tight-binding approximation, the number of Kondo couplings can be reduced to one, for particular impurity configurations.

We have finally calculate, both within large-$N$ expansion technique and renormalization group approach the mean field critical Kondo coupling which is increased by the presence of a renormalized Fermi velocity driven by Coulomb interaction. From the calculation of some thermodynamic quantities, we find, however, that even though the Kondo phase is suppressed, at least in the mean field level, once the singlet is formed, Kondo screening effects are enforced by the Coulomb charge screening. 

\acknowledgments
I would like to thank A. De Martino, D.M. Basko and L. De Leo for useful discussions

\appendix
\section{Comparison with renormalization group}
Redefining the Kondo coupling as $J^\prime=\frac{J\rho }{2\pi\nu}$, with $\nu$ a constant factor, we are going to consider the following generalized Kondo equation, 
\be
\label{dJ}
\frac{d J^\prime}{d \ell}=\frac{d\ln\rho}{d\ell}\,J^\prime+ {J^\prime}^2,
\ee
where $\ell=\ln(D/\epsilon)$ with $\epsilon\ge 0$, a positive defined energy parameter and $\rho$ the density of states, not yet defined. We shall be seeing that, quite in general, 
$-\lim_{\ell\rightarrow \infty}\frac{d\ln\rho}{d\ell}$ 
is the infrared point at criticality which may or may not be a fixed point. 
Solving Eq. (\ref{dJ}) and requiring that $J(D)\equiv J_o$ we obtain
\be
\label{Je}
J^\prime(\epsilon)=\frac{J^\prime_o\,\rho(\epsilon)}{\rho(D)-J^\prime_o\int^D_\epsilon\frac{\rho(x)}{x}dx}.
\ee
For positive coupling, i.e. antiferromagnetic Kondo model, we can define a critical point as
\be
\label{Jc_generic}
J^\prime_c=\frac{\rho(D)}{\int^D_0\frac{\rho(x)}{x}dx},
\ee 
and rewriting Eq. (\ref{Je}) in terms of $J^\prime_c$ we have
\be
\label{Jec}
J^\prime(\epsilon)=\frac{J^\prime_o\,J^\prime_c\,\rho(\epsilon)}{(J^\prime_c-J^\prime_o)\rho(D)+ J^\prime_oJ^\prime_c\int^\epsilon_0\frac{\rho(x)}{x}dx}.
\ee
From these result one can immediately see that, if $J^\prime_o< J^\prime_c$ then $J^\prime(\epsilon)\rightarrow 0$ for $\epsilon \rightarrow 0$ while if $J^\prime_o> J^\prime_c$ then $J^\prime(\epsilon)\rightarrow \infty$ at some low energy scale, called Kondo scale $T_K$. In particular $T_K$ (let us put the Boltzmann constant $k_B=1$) is defined, for $J^\prime_o> J^\prime_c$, by
\be
\label{Tk}
\int^{T_K}_0\frac{\rho(x)}{x}dx=\rho(D)\frac{(J^\prime_o-J^\prime_c)}{J^\prime_oJ^\prime_c}.
\ee
At $J^\prime_o=J^\prime_c$ the infrared limit is given by
\be
\label{lim}
\lim_{\epsilon\rightarrow 0}J^\prime(\epsilon)\Big|_{J^\prime_o=J^\prime_c}=\lim_{\epsilon\rightarrow 0}\frac{\rho(\epsilon)}{\int^\epsilon_0\frac{\rho(x)}{x}dx}=\lim_{\epsilon\rightarrow 0}\frac{\epsilon}{\rho(\epsilon)}\frac{d\rho(\epsilon)}{d\epsilon}=-\lim_{\ell\rightarrow \infty}\frac{d\ln\rho}{d\ell},
\ee
where we have used the de l'H\^opital rule in the second equality.\\
Notice that we get really an unstable infrared fixed point from Eq. (\ref{Jec}) only for a special functional form of the density, let us call it $\bar\rho$, which is the solution of
\be
\bar\rho(\epsilon)=J^\prime_c\int^\epsilon_0\frac{\bar\rho(x)}{x},
\ee
which is $\bar\rho(\epsilon)=\nu \epsilon^{J^\prime_c}$, i.e. a power law. \\
As an example, let us consider logarithmic deviation from power law of the following form
\be
\label{rho}
\rho(\epsilon)=\nu\,\epsilon^r \left[\ln\left(\frac{\Lambda}{\epsilon}\right)\right]^\alpha,
\ee
such that for $r=\alpha=0$ we should get the standard Kondo result and for $r\neq 0$ and $\alpha=0$ we should recover the known results for gapless systems. Assuming the form given by Eq. (\ref{rho}) for the density of states, and using Eq. (\ref{Jc_generic}) we get the following result for the critical value
\be
\label{Jc2}
J^\prime_c=r^{\alpha+1}\frac{D^r}{\Lambda^r} \frac{\left[\ln\left(\frac{\Lambda}{D}\right)\right]^\alpha}{\Gamma\left[1+\alpha,r\ln\left(\frac{\Lambda}{D}\right)\right]},
\ee
which is not universal, dependent (increasingly for $r>0$ and $\alpha<0$) on the ratio $D/\Lambda$.
From Eq. (\ref{Jec}) we have the following expression for the coupling
\be
\label{Je2}
J^\prime(\epsilon)=\frac{r^{\alpha+1} J^\prime_o\,J^\prime_c\,\rho(\epsilon)}{r^{\alpha+1}(J^\prime_c-J^\prime_o)\rho(D)+\nu J^\prime_oJ^\prime_c\Lambda^r\Gamma\left[1+\alpha,r\ln\left(\frac{\Lambda}{\epsilon}\right)\right]}.
\ee
For $J^\prime_o$ smaller than $J^\prime_c$ it flows to zero, while for $J^\prime_o$ greater than $J^\prime_c$ it diverges approaching the following Kondo scale, calculated from Eq. (\ref{Tk}),
\be
T_K\simeq  r^{\frac{\alpha+1}{r}} \rho(D)^{1/r}\frac{(J^\prime_o-J^\prime_c)^{1/r}}{(\nu\, J^\prime_oJ^\prime_c)^{1/r}}\left\{\ln\left[\frac{\Lambda^r \nu\,J^\prime_oJ^\prime_c}{r\rho(D)(J^\prime_o-J^\prime_c)}\right]\right\}^{-\alpha/r}.
\ee  
From Eq. (\ref{Je2}) and using the fact that 
$\lim_{\epsilon\rightarrow 0}\frac{\rho(\epsilon)}{\Gamma\left[1+\alpha,r\ln\left(\frac{\Lambda}{\epsilon}\right)\right]}=\nu\,\Lambda^r r^{-\alpha}$
or, equivalently, using Eq. (\ref{lim}) for $\rho$ given by Eq. (\ref{rho}), we obtain that the infrared limit right at the critical point (i.e. at $J^\prime_o=J^\prime_c$) is 
$\lim_{\epsilon\rightarrow 0}J^\prime(\epsilon)\Big|_{J^\prime_o=J^\prime_c}={r}$, 
which is the same as without logarithmic corrections. We see that this result does not match with $J^\prime_c$ given by Eq. (\ref{Jc2}), except for $\alpha=0$, since, in this case,
$\Gamma\left[1,r\ln\left(\frac{\Lambda}{D}\right)\right]=\frac{D^r}{\Lambda^r}$, 
and we recover the known result $J^\prime_c={r}$ which is then an unstable 
fixed point.

Motivated by a physically relevant example \cite{gonzalez99,polini}, we now use a more realistic density of states induced by the renormalized Fermi velocity Eq.~(\ref{vF}), which then reads
\be
\label{rho_realistic}
\rho(\epsilon)= \frac{\nu \epsilon}{\left(1+\eta \ln(\Lambda/\epsilon)\right)^2}\left(1+\frac{\eta}{1+\eta \ln(\Lambda/\epsilon)}\right).
\ee
Notice that in the infrared limit Eq.~(\ref{rho_realistic}) reduces to Eq.~(\ref{rho}) with $r=1$, $\alpha=-2$ and $\nu\rightarrow \nu/\eta^2$. 
Using Eq.~(\ref{rho_realistic}) in Eq.~(\ref{Jc_generic}) we get the following non-universal critical Kondo coupling
\be
\label{Jc_real}
J^\prime_c=\frac{2\eta^2 v\,\rho(D)}{\eta v_F(D)\rho(D)-v\nu e^{1/\eta}\Lambda \Gamma[0,1/\eta+\log(\Lambda/D)]},
\ee
which is almost equal to Eq.~(\ref{Jc}), after rescaling. It becomes exactly the same if we approximate Eq.~(\ref{rho_realistic}) with $\rho(\epsilon)= \frac{\nu \epsilon}{\left(1+\eta \ln(\Lambda/\epsilon)\right)^2}$, i.e. neglecting the second term.

For $\eta\rightarrow 0$, in the non-interacting limit, we recover $J^\prime_c=1$. Also in the infrared limit $D\rightarrow 0$ we get $J^\prime_c=1$. Moreover, for $D$ going to zero Eq.~(\ref{Jc_real}) approaches Eq.~(\ref{Jc2}) with $r=1$ and $\alpha=-2$, which does not depends on $\eta$. On the other hand, for $D\rightarrow \Lambda$ we have $J^\prime_c= \frac{2\eta^2(1+\eta)}{\eta+\eta^2-e^{1/\eta}\Gamma[0,1/\eta]}$, and for small $\eta$ we get 
\be
J^\prime_c\simeq 1+2\eta.
\ee
Namely, for $D\lesssim \Lambda$, in order to enter the Kondo regime, one needs a starting coupling which exceeds the critical coupling, obtained in the non-interacting case, by approximately an amount directly related to the Thomas-Fermi screening length.


\end{document}